\documentclass[twoside]{article}
\usepackage{fleqn,espcrc2,amssymb}
\usepackage{graphicx}

\newcommand{\AmS}{{\protect\the\textfont2
  A\kern-.1667em\lower.5ex\hbox{M}\kern-.125emS}}
\title{Solution of real-axis Eliashberg equations with different pair symmetries
and tunneling density of states}
\author{G.A.~Ummarino, R.S.~Gonnelli  and D.~Daghero\\
\vspace{3mm} INFM - Dipartimento di Fisica, Politecnico di Torino,
c.so Duca degli Abruzzi 24, 10129 Torino, Italy}
\begin{document}
\begin{abstract}
The real-axis direct solution of the Eliashberg equations for the
retarded electron-boson interaction in the half-filling case and
in the presence of impurities is obtained for six different
symmetries of the order parameter: $s$, $s+\mathrm{i}d$, $s+d$,
$d$, $anisotropic$-$s$ and $extended$-$s$. The spectral function
is assumed to contain an isotropic part
$\alpha_{is}^{2}F\left(\Omega\right) $ and an anisotropic one
$\alpha_{an}^{2}F\left(\Omega\right)$
such that $\alpha_{is}^{2}F\left(\Omega\right)=g\!\cdot\!\alpha%
_{an}^{2}F\left(\Omega%
\right) $, where $g$ is a constant, and the Coulomb
pseudopotential $\mu ^{\ast }$ is set to zero for simplicity. The
density of states is calculated for each symmetry at $T= 2, 4, 40$
and $80$ K. The resulting curves are compared to those obtained by
analytical continuation of the imaginary-axis solution of the
Eliashberg equations and to the experimental tunneling curves of
optimally-doped Bi 2212 crystals. \vspace{-3mm}
\end{abstract}
\maketitle

 In this paper, we make use of the Migdal-Eliashberg
theory \cite{Eliashberg} for the strong electron-boson coupling to
discuss the effect of different possible symmetries of the order
parameter \cite{VanHarlingen} on the tunneling curves of
copper-oxide superconductors. Because of the layered structure of
these materials, we can suppose the quasiparticle wavevectors ${\bf k}$ and $%
{\bf k}^{\prime }$ to lie in the CuO$_{2}$ plane and call $\phi$
and $\phi^{\prime }$ their azimuthal angles in this plane. Then we
solve the Eliashberg equations (EE) using a single-band
approximation with a nearly-circular Fermi line. In the real-axis
formalism the EE take the form of a set of coupled integral
equations for the order parameter $\Delta (\omega ,\phi)$ and the
renormalization function $Z(\omega ,\phi)$,
containing the retarded interaction $\alpha ^{2}(\Omega%
,\phi ,\phi ^{\prime })F(\Omega )$ and the Coulomb pseudopotential
$\mu ^{\ast }\left( \phi,\phi^{\prime }\right) $ \cite {PhysicaC,%
nostro,Rieck}.

We hypothesize that the two last quantities contain an $isotropic$
and an $anisotropic$ part \cite{PhysicaC,nostro} and we expand
both of them in terms of basis functions. Actually, even though we
are able to solve the EE for an arbitrary constant value of
$\mu^{\ast}$, we put it to zero for simplicity. The spectral
function expanded at the lowest order is then expressed by:
$\alpha ^{2}(\Omega,\phi, \phi')F(\Omega)\,$=$\,\alpha _{is}^{2}F(\Omega)\psi _{is}\left( \phi \right)%
\psi_{is}\left( \phi ^{\prime }\right)\,$ +$\, \alpha _{an}^{2}F(\Omega)\psi%
_{an}\left( \phi \right) \psi _{an}\left( \phi ^{\prime }\right)$
where the basis functions $\psi_{is}\left(\phi \right) $ and
$\psi_{an}\left( \phi \right) $ are chosen as follows:
$\psi _{is}\left(\phi\right)$=1; $\psi _{an}\left( \phi \right)$=$\sqrt{2}%
\cos \left(2\phi\right)$ for the \mbox{$d$-wave}, $\psi _{an}\left( \phi \right)$=$8\sqrt{2/35}\cos%
^{4}\left(2\phi\right)$ for the \mbox{$anisotropic$-$s$}, and $\psi%
_{an}\left( \phi \right)$=$-2\sqrt{2/3}\cos^{2}\left(2\phi\right)$
for the \mbox{$extended$-$s$} \cite {VanHarlingen}.  For
simplicity again, we suppose that $\alpha
_{an}^{2}F(\Omega)$=$g\!\!\cdot\!\!\alpha_{is}^{2}F(\Omega)$ where
$g$ is a constant \cite{PhysicaC,nostro}. Thus, the electron-boson
coupling constants for the $isotropic$-wave channel and the
$anisotropic$-wave one, which are given by
$\lambda_{is, an}$= $(1/\pi)\int_{0}^{2\pi}\mathrm{d}\phi\,\psi^{2}_{is, an}(\phi)%
\int_{0}^{+\infty }\mathrm{d}\Omega\, \alpha _{is, an}^{2}F(\Omega
)/\Omega$, result to be proportional:
$\lambda_{an}$=$g\!\cdot\!\lambda_{is}$.

We are interested in solutions of the real-axis EE
of the form: $\Delta (\omega,\phi)$=$\Delta_{is}(\omega%
)$+$\Delta_{an}(\omega)\psi_{an}\left(\phi \right)$ and $Z(\omega%
,\phi)$=$Z_{is}(\omega)$+$Z_{an}(\omega)\psi_{an}\left(\phi
\right)$. The equations are reported explicitly elsewhere
\cite{nostro}. Here we suppose $Z_{an}(\omega )$ to be identically
zero \cite{PhysicaC,nostro}.

The numerical solution of the real-axis EE is performed by using
an iterative procedure. In view of a comparison to the
experimental tunneling curves obtained in Bi~2212 break junctions,
we take $\alpha _{is}^{2}F(\Omega )$=$(\lambda
_{is}/\lambda_{\mathrm{Bi2212}} )\alpha ^{2}F(\Omega
)_{\mathrm{Bi2212}}$ \mbox{\cite{PhysicaC,nostro}}, where $\alpha
^{2}F(\Omega )_{\mathrm{Bi2212}}$ has been experimentally
determined in a previous paper \cite{PhysicaC97} and rescaled to
have $T_{c}$=97 K. Once determined $\Delta (\omega ,\phi )$ and
$Z(\omega ,\phi )$, we calculate the quasiparticle density of
states $N(\omega )$= $\left( 1/2\pi \right) \int_{0}^{2\pi
}\mathrm{d}\phi\,%
\mathrm{Re}\left( \omega / \sqrt{\omega ^{2}- \Delta(\omega ,%
\phi)^{2}}\right) $, whose convolution integral with the Fermi
distribution is the quantity that must be compared to the
experimental tunneling data. Incidentally, the explicit solution
shows that the symmetry of $\Delta (\omega ,\phi )$ is affected by
the choice of the coupling constants $\lambda _{is}$ and $\lambda
_{an}$ and, for some particular values of $\lambda _{is}$ and
$\lambda _{an}$, by the
starting values of $\Delta _{is}(\omega )$ and $ \Delta%
_{an}(\omega )$.

\begin{figure}[t]\vspace{-2mm}
\includegraphics[keepaspectratio,width=7.5cm]{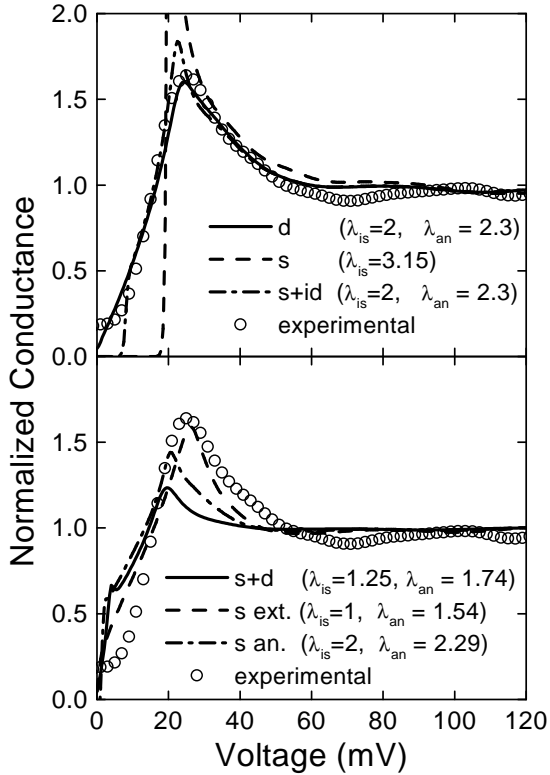}\vspace{-11mm}
\caption{{\small Theoretical tunneling density of states for
various symmetries at $T$=4 K and $\mu ^{\ast}$=0. The circles
represent our experimental data in Bi 2212
\cite{PhysicaC97}.}}\vspace{-7mm}
\end{figure}

The figure reports the theoretical normalized conductance at
$T$=4~K in the six symmetries analyzed and the Bi~2212 experimental %
data obtained in break-junction tunneling experiments \mbox{\cite{PhysicaC,%
PhysicaC97}}. The $d$-wave curve gives the best fit of the peak at
the gap edge and of the conductance behaviour inside the gap,
suggesting that the symmetry of Bi~2212 could be pure $d$-wave.
Nevertheless, none of the symmetries studied here is able to give,
at the same time, $T_{c}$=92-94 K and a peak at 35-41.5 meV, as
experimentally observed by STM on Bi~2212 and recently reported in
literature \cite{Renner}.

The same theoretical curves have also been obtained at $T$=2, 40
and 80~K. The \mbox{$anisotropic$-$s$} and \mbox{$extended$-$s$}
curves are highly temperature-dependent, and some fine structures
evidenced at $T$=2 K are already indistinguishable at 4 K. For
$T\geq T_{c}/3$ it is practically impossible to distinguish
between $d$-wave and mixed symmetries, while the $s$-wave curve
remains clearly distinct.

The curves at $T$=2 K can be compared to those obtained by
analytically continuing the imaginary-axis solutions $\Delta
(\mathrm{i}\omega _{n})$ and $Z(\mathrm{i}\omega _{n})$. In
general, at this low temperature, the analytical continuation
gives a \emph{reasonable agreement} with the real-axis solutions.
The agreement is satisfactory for the $s+\mathrm{i}d$, $d$,
$anisotropic$-$s$ cases, and becomes a little worse in the
$extended$-$s$ case. The imaginary-axis $s$-wave curve is instead
markedly shifted (of about 3 meV) toward higher energies, and then
is unable to approximate the real-axis solution. In the
($s+d$)-wave symmetry, even the shape of the curve is heavily
different in the two cases \cite{nostro}.

Finally, we have studied  the effect of \emph{\mbox{non-magnetic}
impurities} on the tunneling density of states by solving the
appropriate real-axis EE \cite{nostro}. We have found that this
effect is greater on the $d$-wave component of the order
parameter. A small amount of impurities in the unitary limit gives
rise to a zero bias in the $d$-curve, and then improves its
agreement with our experimental points \cite{PhysicaC}. In the
\mbox{non-unitary} case, the $d$-curve peak is further lowered,
broadened and shifted leftward. In the same conditions, all the
mixed-symmetry curves are modified in non-trivial ways, e.g. the
$s+\mathrm{i}d$ and the $anisotropic$-$s$ curves become
practically indistinguishable from the $s$-wave one, but are
shifted toward lower energies. Finally, the $s$-wave curve is
nearly unaffected by this kind of impurities, apart from a small
shift of the gap toward higher energies. \end{document}